\documentstyle[12pt,epsfig]{article}

\def\lapproxeq{\lower .7ex\hbox{$\;\stackrel{\textstyle
<}{\sim}\;$}}
\def\gapproxeq{\lower .7ex\hbox{$\;\stackrel{\textstyle
>}{\sim}\;$}}

\def\lapproxeq{\lower .7ex\hbox{$\;\stackrel{\textstyle
<}{\sim}\;$}}
\def\gapproxeq{\lower .7ex\hbox{$\;\stackrel{\textstyle
>}{\sim}\;$}}

\def\dr{\raisebox{2.1ex}{$\scriptsize\lfloor$}\!\raisebox{1ex}{$\rightarrow$}}
\def\bb{b\bar{b}}

\def\ra{ \rightarrow }

\begin{document}
\titlepage
\begin{flushright}
IPPP/02/41\\
DCPT/02/82\\
3 July 2002
\end{flushright}

\vspace*{2cm}

\begin{center}
{\Large\bf Forward proton tagging as a way to identify a light
Higgs boson at the LHC\footnote{\noindent Presented at the 10th
International Workshop on Deep Inelastic Scattering, DIS(2002),
Krakow, Poland, 30 April--4 May 2002}}

\vspace*{0.5cm}

\textsc{A.D. Martin, V.A. Khoze and M.G. Ryskin}

\vspace*{0.5cm}

{Institute for Particle Physics Phenomenology,\\ University of
Durham, DH1 3LE, UK}
\end{center}

\vspace*{0.5cm}

\begin{abstract}
We show that exclusive double-diffractive Higgs production, $pp\ra
p+H+p$, followed by the $H\ra\bb$ decay, could play an important
role in identifying a `light' Higgs boson at the LHC, provided
that the forward outgoing protons are tagged. We predict the cross
sections for the signal and for all possible $\bb$ backgrounds.
\end{abstract}

\newpage

\section{Introduction}

The identification of the Higgs boson(s) is one of the main goals
of the Large Hadron Collider (LHC) being built at CERN. There are
expectations that there exists a `light' Higgs boson with mass
$M_H\lapproxeq130$~GeV. In this mass range, its detection at the
LHC will be challenging. There is no obvious perfect detection
process, but rather a range of possibilities, none of which is
compelling on its own. Some of the processes are listed in
Table~1, together with the predicted event rates for the
integrated luminosity of 30~fb$^{-1}$ expected over the first two
or three year period of LHC running. We see that, {\em either}
large signals are accompanied by a huge background, {\em or} the
processes have comparable signal and background rates for which
the number of Higgs events is rather small.

Here we wish to draw particular attention to process~(c), which is
often disregarded; that is the exclusive signal $pp\ra p + H + p$,
where the +~sign indicates the presence of a rapidity gap. It is
possible to install proton taggers so that the `missing mass' can
be measured to an accuracy $\Delta M_{\rm missing}\simeq 1$~GeV
\cite{DKMOR}. Then the exclusive process will allow the mass of
the Higgs to be measured in two independent ways. First the tagged
protons give $M_H = M_{\rm missing}$ and second, via the $H\ra\bb$
decay, we have $M_H = M_{\bb}$, although now the resolution is
much poorer with $\Delta M_{\bb}\simeq10$~GeV. The existence of
matching peaks, centered about $M_{\rm missing}=M_{\bb}$, is a
unique feature of the exclusive diffractive Higgs signal. Besides
its obvious value in identifying the Higgs, the mass equality also
plays a key role in reducing background contributions. Another
advantage of the exclusive process $pp\ra p+H+p$, with $H\ra\bb$,
is that the leading order $gg\ra\bb$ background subprocess is
suppressed by a $J_z=0$ selection rule \cite{KMRmm,DKMOR}.

\begin{table}[h]
\begin{center}
\begin{tabular}{l|c|c|c|c}\hline
& \multicolumn{2}{|c|}{number of events}& &  \\
Higgs signal & signal & backgd. & $S/B$ & signif. \\ 
\hline

\raisebox{-3ex}{a) $H\ra\gamma\gamma$}  & \raisebox{-3ex}{313} &
\raisebox{-3ex}{5007} & \raisebox{-3ex}[0pt][7ex]{0.06
${\displaystyle \left(\frac{1\:{\rm GeV}}{\Delta M_{\gamma\gamma}}
\right)}$} & \raisebox{-3ex}{$4.3\sigma$} \\ \hline

\raisebox{-3ex}[0ex][3.3ex]{b) $t\bar{t}H $}  &
\raisebox{-3ex}{26} & \raisebox{-3ex}{31} &
\raisebox{-3ex}[0pt][2ex]{0.8 ${\displaystyle \left(\frac{10\:{\rm
GeV}}{\Delta M_{\bb}} \right)}$} & \raisebox{-3ex}{$3\sigma$} \\   $\qquad\, \dr$ \raisebox{1ex}[0ex][-2ex]{$\bb$}&&&& \\
\hline

\raisebox{-3ex}[0ex][3.3ex]{c) $gg^{PP}\ra p+H+p$} &
\raisebox{-3ex}{11} & \raisebox{-3ex}{4} &
\raisebox{-3ex}[0pt][2ex]{3 ${\displaystyle\left(\frac{1\:{\rm
GeV}}{\Delta M_{\rm
missing}}\right)}$} & \raisebox{-3ex}{$3\sigma$} \\
$\qquad\qquad\qquad\ \ \,\dr$ \raisebox{1ex}[0ex][-2ex]{$\bb$}&&&&\\
\hline

\raisebox{0ex}[3ex][1.9ex]{d) WBF } &  &  &  & \\
\raisebox{0ex}[0ex][3ex]{$qWWq\ra jHj\ra j\tau\tau j$} & 25 & 8 & 3 & $4.4\sigma$ \\
\hline

\raisebox{0ex}[3ex][1.9ex]{e) WBF with rap. gaps } &  &  & \\
{$qWWq\ra j+H+j$} & {250} & {1800} &
\raisebox{0ex}[2ex][1.5ex]{0.14\ {$\displaystyle
\left(\frac{10\:{\rm GeV}}{\Delta
M_{\bb}} \right)$}}  & {$5.5\sigma$} \\
$\qquad\qquad\qquad\ \; $\raisebox{-0.5ex}[0pt][2ex]{$\dr$}
\raisebox{0.5ex}[0pt][2ex]{$\bb$}&  &  & \\ \hline

\end{tabular}
\end{center}
\begin{caption}
{\footnotesize The number of signal and background events for
various methods of Higgs detection at the LHC. The significance of
the signal, $S/\sqrt{S+B}$, is also given. The mass of the Higgs
boson is taken to be 120~GeV and the integrated luminosity is
taken to be 30 fb$^{-1}$. The notation $gg^{PP}$ is to indicate
that the gluons originate within overall colour-singlet (hard
Pomeron) $t$-channel exchanges. The entries for the various
processes are computed from references (a) \cite{Z}, (b)
\cite{TT}, (c) \cite{DKMOR,INC}, (d) \cite{Z,WBF} and (e)
\cite{DKMOR,KMRhiggs}. For~(a) we show the CMS value without the
NLO $K$ factor. Including the $K$ factors for both the signal and
the background increases the $H\ra\gamma\gamma$ significance to
about $7\sigma$~\cite{BDS}.}
\end{caption}
\end{table}

\section{Calculation of the exclusive Higgs signal}

The basic mechanism for the exclusive process, $pp\ra p+H+p$, is
shown in Fig.~1. Since the dominant contribution comes from the
region $\Lambda_{\rm QCD}^2\ll Q_t^2\ll M_H^2$ the amplitude may
be calculated using perturbative QCD techniques \cite{KMR,KMRmm}
\begin{equation}
{\cal M} = \frac{(\sqrt{2}G_F)^{\frac{1}{2}}\pi^2\alpha_S}{3} \int
\frac{d^2Q_t}{Q_t^4} f_g\left(x_1,x_1^\prime, Q_t^2,
\frac{M_H^2}{4}\right) f_g\left(x_2,x_2^\prime, Q_t^2,
\frac{M_H^2}{4}\right), \label{eq:M}
\end{equation}
where the skewed unintegrated gluon densities, $f_g$, are given in
terms of the conventional integrated density $g(x)$. The $f_g$'s
embody a Sudakov suppression factor $T$, which is effectively the
survival probability that the gluon remains untouched in the
evolution from $Q_t$ up to the hard scale $M_H/2$.
\begin{center}
\begin{figure}[h]
\epsfig{figure=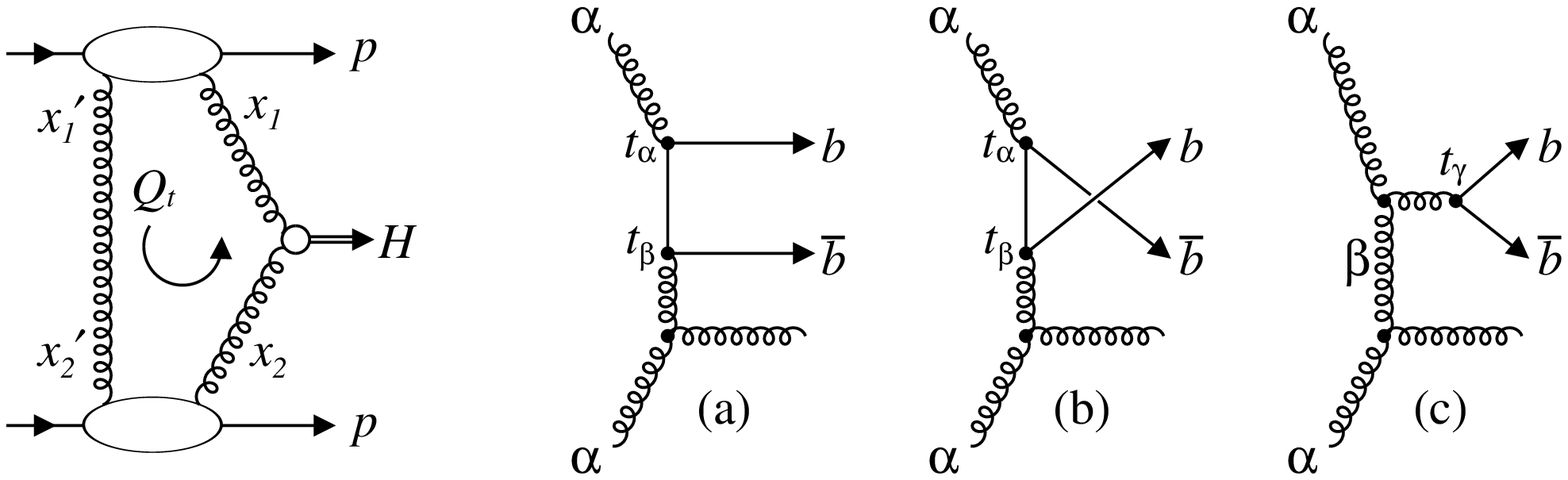,width=13.6cm}
 \label{Fig.1}
\end{figure}
\end{center}
\vspace{-1cm}
\begin{tabular}{p{3.5cm} p{0.4cm} p{8.5cm}}
\footnotesize{Fig.1:~Schematic diagram for the exclusive process
$pp\ra p+ H + p$.} & & \footnotesize{Fig.2:~The background from
colour-singlet NLO $gg\ra\bb g$ production, where $\alpha$,
$\beta$ and $\gamma$ are gluon colour labels and where the $t_i$
are the colour matrices for the quark-gluon vertices.}
\end{tabular}

\vspace{2ex} The radiation associated with the $gg\ra H$ hard
subprocess is not the only way to populate and to destroy the
rapidity gaps. There is also the possibility of soft rescattering
in which particles from the underlying event populate the gaps.
The probability, $S^2=0.02$, that the gaps survive the soft
rescattering was calculated using a two-channel eikonal model,
which incorporates high mass diffraction \cite{KMRsoft}. Including
this factor, and the NLO $K$ factor, the cross section is
predicted to be \cite{INC}
\begin{equation}
\sigma(pp\ra p+H+p)\simeq 3\:{\rm fb} \label{eq:sigma}
\end{equation}
for the production of a Standard Model Higgs boson of mass 120~GeV
at the LHC\footnote{Cross section (\ref{eq:sigma}) at the
Tevatron, 0.2~fb, is too low to provide a viable signal.}. It is
estimated that there may be a factor two uncertainty in this
prediction \cite{DKMOR}.

The event rate in entry~(c) of Table~1 includes a factor 0.6 for
the efficiency associated with proton tagging, 0.6 for $b$ and
$\bar{b}$ tagging, 0.5 for the $b,\bar{b}$ jet polar angle cut,
$60^\circ<\theta<120^\circ$, (necessary to reduce the $\bb$ QCD
background) and 0.67 for the $H\ra\bb$ branching fraction
\cite{DKMOR}. Hence the original $(\sigma=3\:{\rm fb})\times({\cal
L}=30\:{\rm fb}^{-1}) = 90$ events is reduced to an observable
signal of 11 events, as shown in Table~1.

\section{Background to the exclusive Higgs signal}

The advantage of the $p+(H\ra\bb)+p\,$ signal is that there exists
a $J_z=0$ selection rule, which requires the leading order
$gg^{PP}\ra\bb$ background subprocess to vanish in the limit of
massless quarks and forward outgoing protons\footnote{In the
$m_b\ra0$ limit, the two Born-level diagrams (Figs.~2(a,b) {\em
without} the emission of the gluon) cancel each other.}. However,
in practice, LO background contributions remain. The prolific
$gg^{PP}\ra gg$ subprocess may mimic $\bb$ production since we may
misidentify the outgoing gluons as $b$ and $\bar{b}$ jets.
Assuming the expected 1\% probability of misidentification, and
applying $60^\circ<\theta<120^\circ$ jet cut, gives a
background-to-signal ratio $B/S \sim 0.06$. Secondly, there is an
admixture of $|J_z|=2$ production, arising from non-forward going
protons which gives $B/S \sim 0.08$. Thirdly, for a massive quark
there is a contribution to the $J_z=0$ cross section of order
$m_b^2/E_T^2$, leading to $B/S \sim 0.06$, where $E_T$ is the
transverse energy of the $b$ and $\bar{b}$ jets.

Next, we have the possibility of NLO $gg^{PP}\ra\bb g$ background
contributions. Of course, the extra gluon may be observed
experimentally and these background events eliminated. However,
there are exceptions. The extra gluon may go unobserved in the
direction of a forward proton. This background may be effectively
eliminated by requiring the equality $M_{\rm missing} = M_{\bb}$.
Then we may have soft gluon emission. First, we note that emission
from an outgoing $b$ or $\bar{b}$ is not a problem, since we
retain the cancellation between the crossed and uncrossed graphs.
Emission from the virtual $b$ line is suppressed by at least a
factor of $\omega/E$ (in the amplitude), where $\omega$ and $E$
are the energies of the outgoing soft gluon and an outgoing $b$
quark in the $gg^{PP}\ra\bb$ centre-of-mass frame. The potential
danger is gluon emission from an incoming gluon, see Fig.~2. The
first two diagrams no longer cancel, as the $\bb$ system is in a
colour-octet state. However, the third diagram has precisely the
colour and spin structure to restore the cancellation. Thus soft
gluon emissions from the initial colour-singlet $gg^{PP}$ state
factorize and, due to the overriding $J_z=0$ selection rule, QCD
$\bb$ production is still suppressed. The remaining danger is
large angle hard gluon emission which is collinear with either the
$b$ or $\bar{b}$ jet, and therefore unobservable. If the cone
angle needed to separate the $g$ jet from the $b$ (or $\bar{b}$)
jet is $\Delta R \sim 0.5$ then the expected background from
unresolved three jet events leads to $B/S \simeq 0.06$.

The NNLO $\bb gg$ background contributions are found to be
negligible (after requiring $M_{\rm missing}\simeq M_{\bb}$), as
are soft Pomeron-Pomeron fusion contributions to the background
(and to the signal)~\cite{DKMOR}. So, in total, double-diffractive
Higgs production has a signal-to-background ratio of about three,
after including the $K$ factors.

\section{Discussion}

Identifying a `light' Higgs will be a considerable experimental
challenge. All detection processes should be considered. From
Table~1 we see that valuable information can be obtained from weak
boson fusion, where the Higgs and the accompanying jets are
produced at high $p_t$. For example, process~(d) is based on the
$H\ra\tau\tau$ decay for which the background is small
\cite{Z,WBF}, whereas process~(f) exploits rapidity gaps so that
the larger $H\ra\bb$ signal may be isolated \cite{KMRhiggs},
provided the pile-up problems can be overcome \cite{DKMOR}.

Here we have drawn attention to the exclusive $pp\ra p+H+p$
signal, process~(c). The process has the advantage that the signal
exceeds the background. The favourable signal-to-background ratio
is offset by a low event rate, caused by the necessity to preserve
the rapidity gaps so as to ensure an exclusive signal.
Nevertheless, entry~(c) of Table~1 shows that the signal has
reasonable significance in comparison to the standard
$H\ra\gamma\gamma$ and $t\bar{t}H$ search modes. Moreover, the
advantage of the matching Higgs peaks, $M_{\rm missing} =
M_{\bb}$, cannot be overemphasized \footnote{This may be
contrasted with the search for a Higgs peak sitting on a huge
background in the $M_{\gamma\gamma}$ spectrum, see process~(a) of
Table~1.}.

\section*{Acknowledgements}
We thank Albert De Roeck, Risto Orava and Andrei Shuvaev for
valuable discussions, and the EU, PPARC and the Leverhulme Trust
for support.

\end{document}